\documentclass[prd,amsmath,showpacs,preprintnumbers,
               floatfix,twocolumn,letterpaper,superscriptaddress,
h-physrev4-1]{revtex4-1}
\usepackage{times}
\usepackage{graphicx}
\usepackage{slashed}
\usepackage{amssymb}
\usepackage{mathrsfs}
\newcommand{\nc}{\newcommand}
\nc{\lb}{\langle}
\nc{\rk}{\rangle}
\nc{\Blb}{\Big\langle}
\nc{\Brk}{\Big\rangle}

\nc{\mi}{\!\!\mid\!\!}
\nc{\ra}{\rightarrow}
\nc{\Ra}{\Rightarrow}
\nc {\cd}{\partial}
\nc {\sla}{\slashed}
\nc{\ro}{\mathrm}
\nc{\ca}{\mathcal}
\nc{\sr}{\mathscr}
\nc{\bo}{\mathbf}
\nc{\Tr}{\ro{Tr}\,}
\nc{\Str}{\ro{Str}}
\nc{\realtrace}{\ro{Re\; Tr}}
\nc{\maxrealtrace}{\ro{max\, Re\; Tr}}
\nc{\ud}{\ro{d}}
\nc{\nn}{\nonumber}

\nc{\pb}{\bar{\psi}}
\nc{\p}{\psi}
\nc{\Pb}{\bar{\Psi}}
\nc{\vp}{\vec{\pi}}
\nc{\vap}{\varphi}
\nc{\vt}{\vec{\tau}}
\nc{\si}{\sigma}
\nc{\Si}{\Sigma}
\nc{\tSi}{\tilde{\Sigma}}
\nc{\g}{\gamma}
\nc{\G}{\Gamma}
\nc{\la}{\lambda}
\nc{\La}{\Lambda}
\nc{\ep}{\epsilon}
\nc{\de}{\delta}
\nc{\De}{\Delta}
\nc{\cL}{\ca{L}}
\nc{\cLe}{\ca{L}_{\ro{eff}}}
\nc {\ti}{\tilde}
\nc{\f}{\frac}
\nc{\da}{\dagger}
\nc{\SU}{\ro{SU}}
\nc{\om}{\omega}
\nc{\Om}{\Omega}

\nc{\darrow}{\stackrel{\leftrightarrow}{\cd}}
\nc{\darrows}{\stackrel{\leftrightarrow}{\sla{\cd}}}
\nc{\Darrows}{\stackrel{\leftrightarrow}{\sla{D}}}
\nc{\mr}{\stackrel{\circ}{m}_\rho}

\nc {\eqb}{\begin{equation}}
\nc {\eqe}{\end{equation}}
\nc {\eqab}{\begin{eqnarray}}
\nc {\eqae}{\end{eqnarray}}

\begin{document}

%(Augmented/Extended)
%\title{The intrinsic scale of the nucleon: magnetic moments in extended effective field theory}

\title{Chiral extrapolations for nucleon magnetic moments}

\author{J. M. M. Hall} 
\affiliation{Special Research Centre for the Subatomic Structure of
  Matter (CSSM), School of Chemistry and Physics, University of Adelaide 5005,
  Australia}

\author{D. B. Leinweber}
\affiliation{Special Research Centre for the Subatomic Structure of
  Matter (CSSM), School of Chemistry and Physics, University of Adelaide 5005,
  Australia}

\author{R. D. Young} 
\affiliation{Special Research Centre for the Subatomic Structure of Matter 
  (CSSM), School of Chemistry and Physics, University of Adelaide 5005,
  Australia}
\affiliation{ARC Centre of Excellence for Particle Physics at the Terascale, 
School of Chemistry and Physics, University of Adelaide 5005,
  Australia}

%New preprint no.
%\preprint{ADP-12-46/T768}
%CORRECTED:
\preprint{ADP-12-01/T768}

%ABSTRACT
\begin{abstract}

  Lattice QCD simulations have made
  significant progress in the calculation of nucleon electromagnetic
  form factors in the chiral regime in recent years. 
With simulation results achieving 
  pion masses of order $\sim 180\,\,{\rm MeV}$, there is an apparent
  challenge as to how the physical regime is approached. By using
  contemporary methods in chiral effective field theory ($\chi$EFT), both the
  quark-mass and finite-volume dependence of the isovector nucleon 
  magnetic moment are carefully examined. The extrapolation to the
  physical point yields a result that is compatible with experiment,
  albeit with a combined statistical and systematic uncertainty of
  $\pm 10\%$. The extrapolation shows a strong finite-volume dependence;  
  lattice sizes of $L>5\,\,{\rm fm}$
  must be used to simulate results within $2\%$ of the infinite-volume 
result for the magnetic
  moment at the physical pion mass. 
  %muphysinf-2% = 4.2316
  %muphys(L=5.1fm) = 4.2333

\end{abstract}

\maketitle

%*****************************************%
\section{Introduction}
\label{sect:intro}

%The analysis of the magnetic moment of the nucleon provides an excellent 
%check for the correct identification of an intrinsic scale in the nucleon-pion 
%interaction. %Using chiral effective field theory ($\chi$EFT), 
%It has been demonstrated previously that such an intrinsic scale 
%may be extracted from the results of lattice quantum chromodynamics (QCD) 
%calculations for the mass of the nucleon \cite{Hall:2010ai}. 
%%that lattice quantum chromodynamics (QCD) results for the nucleon mass 
%%have an energy scale embedded within them \cite{Hall:2010ai}. 
%This property is %intimately related to 
% a consequence of the size of the %existence of a
% power-counting regime (PCR), defined where the expansion formulae  
%of chiral perturbation theory ($\chi$PT), to finite chiral order, 
%formally hold. Since a selection 
%of lattice QCD results reasonable for fitting an extrapolation invariably 
%extend outside the restrictive PCR \cite{Leinweber:2005xz}, 
%the validity of a formal scheme 
%for extrapolation, and for identifying the leading-order terms in the 
%chiral expansion, is %jeopardized.
%compromised. 

%The extrapolation of lattice QCD results to the physical regime 
%presents an 

%define QCD, $\chi$EFT
The distribution of the electric and magnetic charge currents of the
nucleon are characterized by the elastic electromagnetic form factors
--- for recent reviews on experimental progress, see
Refs.~\cite{Arrington:2011kb,Perdrisat:2006hj,Arrington:2006zm,HydeWright:2004gh,Gao:2003ag}.
The description of the electromagnetic form factors in terms of the
elementary degrees of freedom of QCD has seen significant progress
through recent advances in lattice QCD simulations
\cite{Yamazaki:2009zq,Syritsyn:2009mx,Bratt:2010jn,Alexandrou:2011db,Collins:2011mk}.
%The development of lattice quantum chromodynamics (QCD) as a tool for
%the investigation of electromagnetic form factors has contributed much
%to the current understanding of magnetic moments and charge
%distributions within hadrons.

Lattice QCD simulations of the electromagnetic form factors 
of the nucleon \cite{Leinweber:1990dv} are now probing into the chiral regime, 
where the QCDSF Collaboration have recently reported results at pion masses 
as low as  %$m_\pi 
$\sim 180\,\,{\rm MeV}$ \cite{Collins:2011mk}.
%generated by the 
The results of this work have presented a challenge in the pion-mass
extrapolation to the physical point. The results presented in
Ref.~\cite{Collins:2011mk} have been used to investigate 
the applicability of a range
of chiral effective field theory ($\chi$EFT) methods, including
``heavy-baryon'' \cite{Bernard:1992qa}, ``small scale expansion''
\cite{Hemmert:2002uh}, and ``covariant baryon'' approaches
\cite{Dorati:2007bk}. %Ref.~\cite{Collins:2011mk} 
It has been  demonstrated that there is a
difficulty in achieving a consistent quantitative description of the
pion-mass dependence of key observables (such as magnetic moments
and charge radii) between the physical point and the lightest
simulation results \cite{Collins:2011mk}. This issue has persisted 
across a number of simulations \cite{Gockeler:2003ay,Boinepalli:2006xd}.
In the present manuscript, a chiral extrapolation
of the QCDSF results is developed for the isovector nucleon magnetic moment
based on finite-range regularized (FRR) $\chi$EFT
\cite{Leinweber:2003dg,Young:2004tb}.

%The magnetic moment is of interest because of the physical significance 
%of its anomalous component, obtained from the Pauli form factor. Because 
% electrically charged pions with non-zero angular momentum
% dress the bare nucleon, 
%they contribute non-trivially to its magnetic moment, 
%altering the value from its semi-classical Dirac value. 
In the application of FRR to the extrapolation of the magnetic moment,
new developments are utilized in order to ensure the robustness of the
extrapolation procedure. These include the identification of the
preferred finite regularization scale directly from the lattice results
\cite{Hall:2010ai} and a determination of an upper bound of the pion
mass that can be reliably incorporated in the extrapolation
\cite{Hall:2011en}. A feature of the analysis is that the combination
of both finite-volume corrections and the onset of rapid nonanalytic
behaviour in the chiral regime leads to an extrapolation that is
compatible with the experimental value. The analysis provides the predicted
quark-mass dependence for a range of fixed-volume lattices, which act
to emphasize the importance of achieving large volumes in order to
reveal the strong nonanalytic behaviour directly on the lattice.

\begin{figure}[bp]
\includegraphics[height=0.950\hsize,angle=90]{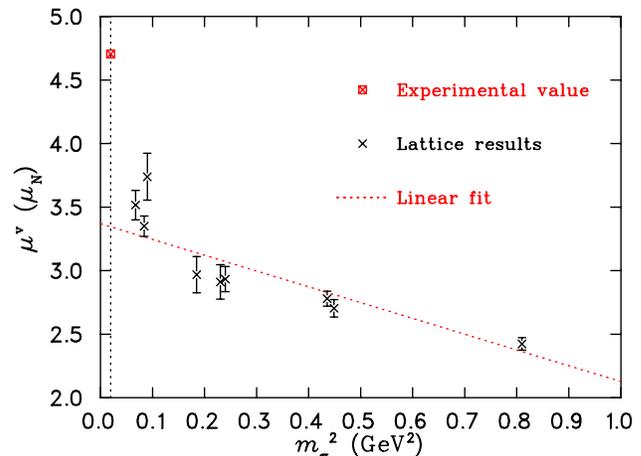}
\vspace{-11pt}
\caption{\footnotesize{(color online). Lattice QCD data for $\mu_N^\ro{v}$ 
from QCDSF \cite{Collins:2011mk}, with experimental value as marked \cite{Mohr:2008fa,Nakamura:2010zzi}. The lattice results satisfy $L>1.5$ fm, $m_\pi L>3$. A simple linear
 fit is also included, which misses the experimental value.}}
\label{fig:data}
\end{figure}

%Recent lattice QCD simulations by 
%the QCDSF collaboration %results for 
% have determined the magnetic moment of the isovector nucleon, 
%using $N_f = 2$ and 
%$\ca{O}(a)$-improved
% Wilson quark action \cite{Collins:2011mk} . %at a variety of $m_\pi^2$ values 
 The lattice QCD results for the magnetic moment of 
the isovector nucleon, $\mu_N^\ro{v}$,  from the QCDSF Collaboration 
are displayed in Fig. \ref{fig:data}. 
The lattice calculation used $N_f = 2$ and 
the $\ca{O}(a)$-improved Wilson quark action \cite{Collins:2011mk}.
%In lattice QCD 
The isovector combination $(p - n)$ 
is considered to avoid calculating 
the disconnected loops that occur in full QCD. 
%The 
%lattice results are obtained from Ref.~\cite{Collins:2011mk}, 
%and 
%Isovector nucleon results for two-flavour $\ca{O}(a)$-improved
% Wilson quark action 
%from the QCDSF Collaboration \cite{Collins:2011mk} 
%is used in this analysis, which is displayed 
%in Fig. \ref{fig:data}, with the restrictions 
%$L>1.5$ fm and $m_\pi L > 3$. There are nine lattice points that satisfy these 
%criteria from the original set of results. 
%
To ensure that the lattice results %are not too far from 
give a reasonable approximation to the infinite-volume limit, 
the following restrictions are applied: $L>1.5$ fm and $m_\pi L > 3$. 
There are nine lattice points that satisfy these 
criteria from the original set of results.  
 The lattice sizes %of each 
%data point in this reduced set 
 considered 
vary from $1.7$ fm to $2.9$ fm. 
A simple linear fit is included in this plot, which does not take into account 
the chiral loop integrals, nor the finite-volume corrections to the data.
%Therefore, 
Neglecting these effects, it is not surprising that the 
 linear trend does not reach the experimental value of the 
magnetic moment at the physical pion mass.
%Since the lattice QCD results extend outside the PCR, the result of 
%an extrapolation will be regulator-dependent. However, scheme-dependence 
%can be handled by obtaining an optimal regularization scale 
%in the extrapolations. 
%as described in Ref. \cite{Hall:2010ai}.
The use of extended $\chi$EFT methods in performing the extrapolation to 
the physical point will now be explored.

\section{Chiral effective field theory}
\label{sect:eft}

%In an effective field theory of the electromagnetic properties %observables 
%of baryons, such as the nucleon, 
The elastic 
matrix element for the baryon-photon interaction can be parametrized 
by the Dirac and Pauli form factors, $F_1$ and $F_2$, respectively,  
%In the non-relativistic limit, this can be 
written as:
\begin{align}
& \lb B(p')\mi J_\mu\mi B(p)\rk = \nn\\
&\quad \bar{u}^{s'}(p')\left\{\g_\mu\, F_1(Q^2)
+\f{i\si_{\mu\nu} q^\nu}{2m_B}\,F_2(Q^2)\right\}u^s(p)\,.
\end{align}
$Q^2$ is a positive momentum transfer $Q^2 = -(p'-p)^2$. The Sachs 
electromagnetic form factors $G_{E,M}$ are the linear combinations 
of $F_1$ and $F_2$ defined 
by:
\begin{align}
G_E(Q^2) &= F_1(Q^2) - \f{Q^2}{4m_B^2}F_2(Q^2)\,,\\
G_M(Q^2) &= F_1(Q^2) + F_2(Q^2)\,.
\end{align}
%
%By considering the behavior of 
The Sachs magnetic form factor 
of the nucleon at zero-momentum 
transfer, $G_M(Q^2 = 0)$, %, one obtains 
defines the magnetic moment as two separate 
terms: %an anomalous component due to the finite-size of  
%hadron interactions, and unity due to charge conservation (in physical units):
 the Dirac moment (unity), plus an anomalous contribution associated 
with the internal structure of the hadron:
\begin{align}
\label{eqn:nucmagmom}
\mu_N^\ro{v} &= G_{M}^\ro{v}(Q^2=0) \\
&= 1 + \kappa_n.
\end{align}

%For octet baryons, the magnetic moments obey the Coleman-Glashow $\SU(3)$ 
%relations related to the following Langrangian of two independent terms \cite{Jenkins:1992pi,Wang:2007iw,Wang:2008vb,Wang:1900ta}:
%
%\begin{align}
%\label{eqn:lag}
%\cL_{\ro{oct}}^{\ro{e-m}} &= \f{e}{4m_N}\Big(\mu_D\,\Tr \bar{B}_\ro{v} \si^{\mu\nu}
%\{F_{\mu\nu}^+,B_\ro{v}\}\nn\\
%&\qquad\quad\,+\mu_F\,\Tr \bar{B}_\ro{v}\si^{\mu\nu}[F_{\mu\nu}^+,B_\ro{v}]\Big)\,.
%\end{align}
%
%For an Abelian gauge field $\sr{A}_\mu$ with field strength tensor 
%$F_{\mu\nu}\equiv \cd_{[\mu}\sr{A}_{\nu]}$, and $\SU(3)$ quark charge 
%matrix $\ca{Q} = \ro{diag}(2/3,-1/3,-1/3)$, 
%the quantity $F_{\mu\nu}^+$ has been chosen such that it is invariant under local 
%chiral symmetry transformations:
%
%\eqb
%F_{\mu\nu}^+ \equiv \f{1}{2}(\xi^\da F_{\mu\nu}\ca{Q}\xi 
%+ \xi F_{\mu\nu}\ca{Q}\xi^\da)\,,
%\eqe
%

For the leading-order contributions to the magnetic moment, the 
standard first-order interaction 
Lagrangian from heavy-baryon chiral perturbation theory ($\chi$PT) is used 
\cite{Jenkins:1991ne,Jenkins:1990jv,Jenkins:1991ts,Labrenz:1996jy,WalkerLoud:2004hf,Wang:2007iw}:
\begin{align}
\label{eqn:lag}
\cL^{(1)}_{\chi PT}  &= 2D\,\Tr[\bar{B}_\ro{v}S_\ro{v}^\mu\{A_\mu,B_\ro{v}\}\,]
+2F\,\Tr[\bar{B}_\ro{v}S_\ro{v}^\mu[A_\mu,B_\ro{v}]\,] \nn \\
&+\ca{C}\,(\bar{T}_\ro{v}^\mu A_\mu B_\ro{v} +
 \bar{B}_\ro{v}A_\mu  T_\ro{v}^\mu),\\
(S_\ro{v}^\mu &= \f{i}{2}\gamma_5 \si^{\mu\nu}v_\nu),
\end{align}
where the pseudo-Goldstone 
fields are encoded as the adjoint representation of 
$\SU(3)_L\otimes\SU(3)_R$, forming an axial vector combination $A_\mu$:
\begin{align}
\xi &\equiv \ro{exp}\left\{{\f{i}{f_\pi}\tau^a \pi^a}\right\},\\
A_\mu &= \f{1}{2}(\xi\,\cd_\mu\,\xi^\da - \xi^\da\,\cd_\mu\,\xi).
\end{align}
By the convention presented here, $f_\pi = 92.4$ MeV. 
The values for the couplings in 
the interaction Lagrangian are obtained from the 
 $\SU(6)$
 flavour-symmetry relations \cite{Jenkins:1991ts,Lebed:1994ga} 
and from phenomenology: $D = 0.76$, $F = \f{2}{3}D$
and $\ca{C} = -2D$.

\begin{figure}[bp]
\centering
\includegraphics[height=80pt]{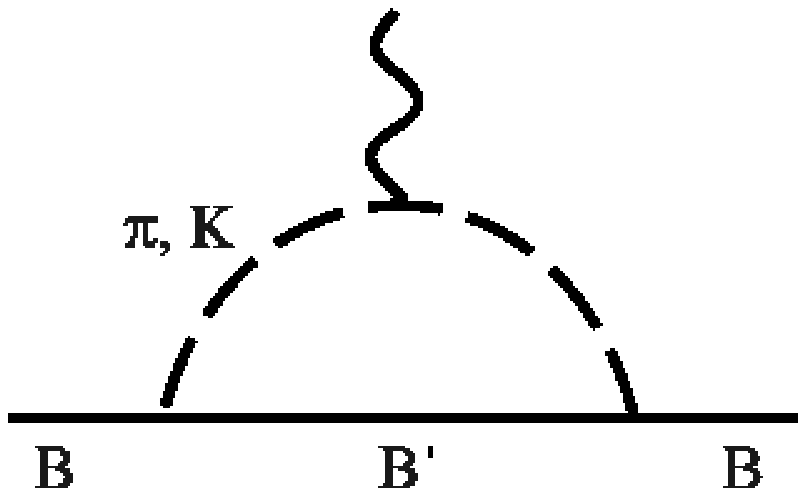}
\caption{\footnotesize{The pion/kaon loop contributions to the magnetic moment 
of 
an octet baryon $B$, allowing a transition to a baryon $B'$, 
with a photon attachment, 
 which provides the leading nonanalytic contribution. % to
%    the magnetic moment. 
All charge conserving transitions are implicit.}}
\label{fig:emSEa}
%\end{figure}
%
%\begin{figure}
\centering
\includegraphics[height=80pt,angle=0]{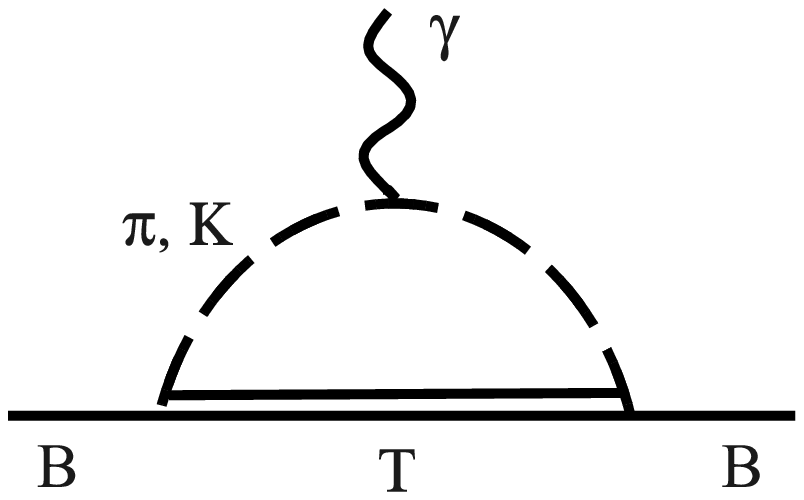}
\caption{\footnotesize{The pion/kaon loop contribution to the magnetic moment 
    of an octet baryon $B$,  
allowing transitions to nearby and
    strongly coupled decuplet baryons $T$.}}
\label{fig:emSEb}
%\end{figure}
%
%\begin{figure}
%\centering
%\includegraphics[height=80pt,angle=0]{emSEc.eps}
%\caption{\footnotesize{Tadpole contributions at $\ca{O}(m_q)$ to the octet 
%baryon self energy, with photon attachment.}}
%\label{fig:emSEc}
\end{figure}

From the full Lagrangian, the chiral behaviour of the magnetic moment can 
be written in terms of an ordered expansion in pion mass squared, 
through use of the Gell-Mann$-$Oakes$-$Renner
Relation, $m_q \propto m_\pi^2$ \cite{GellMann:1968rz}:
\eqb
\label{eqn:chiral}
\mu_N^\ro{v} = a_0^\La + a_2^\La\,m_\pi^2 + \ca{T}_N(m_\pi^2\,;\La) + 
\ca{T}_\De(m_\pi^2\,;\La) + 
 \ca{O}(m_\pi^4)\,.
\eqe
This expansion contains an analytic polynomial in $m_\pi^2$ plus 
the leading-order chiral loop integrals ($\ca{T}_{N,\De}$), 
from which nonanalytic behaviour arises.
The coefficients $a_i^\La$ are the (scale-dependent)  
`residual series' coefficients. %, which 
%correspond to direct quark-mass insertions in the full
%Lagrangian. 
Upon renormalization of the divergent loop 
integrals, these terms correspond to low-energy coefficients of $\chi$EFT 
\cite{Young:2002ib}. 
In this instance, only two free parameters are provided in the residual series. 
% since the 
%nonanalytic contributions are included only to order 
%$\ca{O}(m_\pi^4\,\ro{log}\,m_\pi)$. 
The leading-order diagrams included in 
this investigation are simply the $1$-meson  loops, %in $N_f=2$ QCD, 
as shown in Figs. \ref{fig:emSEa} and \ref{fig:emSEb}. 

\subsection{Finite-range regularization}
\label{subsect:frr}

Finite-range regularized effective field theory handles divergences 
in the ultraviolet region of the loop integrals by introducing a 
regulator function as part of the coupling to each vertex of the diagram. 
The regulator function $u(k\,;\La)$ introduces a cutoff scale $\La$, 
and should satisfy %$u(k=0;\La)=1$ and $u(k\rightarrow\infty;\La)=0$.
$u|_{k=0} = 1$ and $u|_{k\rightarrow\infty} = 0$.
The exact functional form chosen for the regulator is  
independent of the result of calculation, so long as the lattice QCD results 
are constrained within the power-counting regime (PCR). 
%In order to avoid obfuscating the 
%calculation with insignificant 
%regularization-dependent %inconvenient 
%finite-volume %artifacts, as is the case
%corrections, 
%for the sharp cutoff regulator form,  
A smooth attenuating dipole form is chosen for this investigation:
\eqb
u(k\,;\La) = {\left(1 + \f{k^2}{\La^2}\right)}^{-2}.
\eqe
Detailed analyses exist for a range of alternative forms 
\cite{Leinweber:2005xz,Hall:2010ai}.
Though it has been suggested that a 
sharp cutoff FRR scheme should be chosen to ensure the 
preservation of the chiral %Slavnov-Taylor 
Ward Identities
 \cite{Bernard:2003rp}, it is possible to maintain chiral symmetry 
 by including the necessary vertex corrections 
at higher-order in order to suppress any extra scheme-dependent 
nonanalytic terms induced by regulators such as the dipole 
\cite{Djukanovic:2004px}. Here, chiral Ward Identities are maintained 
to the order of the calculation. 
%
%However, since the renormalization of 
% a single observable at any time is considered in this analysis, 
%%namely, the magnetic moment, 
% the issue is not so urgent. 

%problem: we are comparing the intrinsic scale result of the 
%nucleon mass to the result 
%of the magnetic moment? Do we need to ensure chiral Slavnov-Identities 
%are respected?

FRR, in conjunction 
with $\chi$EFT-inspired techniques, 
provides a robust method for achieving an extrapolation 
to physical quark masses, and identifying an intrinsic scale embedded 
within lattice QCD results. 
It has been demonstrated previously that such an intrinsic scale 
may be extracted from the results of lattice QCD 
calculations for the mass of the nucleon \cite{Hall:2010ai}. 
This property is 
 a consequence of the size of the %existence of a
 PCR, defined where the expansion formulae  
of $\chi$PT formally hold (to finite chiral order). 
This %so-called 
%\emph{extended} 
extended effective field 
theory %(E$^2$FT) 
proceeds by analyzing the behaviour of the renormalization 
of one or more low-energy coefficients of the chiral expansion as a
function of the regularization scale. Ideally, with lattice QCD 
results constrained entirely within the PCR, the renormalized 
coefficients are independent of regularization scale. However, 
in practice, a scale-dependence is observed; %curvature; 
particularly for lattice result sets including %involving 
points corresponding to quark masses beyond the PCR. 
%By analyzing a variety of subsets of the lattice QCD results %for varying 
%quark masses, 
By truncating the lattice QCD results at different quark-mass 
points (corresponding 
to a value of $m_{\pi,\ro{max}}^2$), 
an optimal regularization scale can be identified. 
This optimal scale is the value at which the low-energy coefficients 
are least sensitive to 
 the truncation of the lattice results \cite{Hall:2010ai}. 

\subsection{Loop integrals and definitions}
\label{subsect:loop}

The value of the magnetic moment is renormalized by contributions from 
loop integrals, obtained from the effective field theory. Here, the focus 
is on the pion contributions, noting that it is straightforward to include 
the kaon contribution, using $m_K^2 = m_{K,\ro{phys}}^2 + \f{1}{2}(m_\pi^2
-m_{\pi,\ro{phys}}^2)$, as is done in Sec. \ref{sect:results} reporting 
the results. 
The two leading-order loop integrals are the $1$-meson loops, as shown 
in Figs. \ref{fig:emSEa} and \ref{fig:emSEb}. 
Each loop integral can be expanded as a polynomial series, which is analytic 
in quark mass,  
%has a solution in the form of a polynomial expansion
%analytic in $m_\pi^2$ 
plus a nonanalytic term:
\begin{align}
\label{eqn:Sia}
\ca{T}_N(m_\pi^2\,;\La) 
&=  b_0^{N} + \chi_N\,m_\pi + b_2^{N}\,m_\pi^2 + \ca{O}(m_\pi^3)\,,\\
\label{eqn:Sib}
\ca{T}_\De(m_\pi^2\,;\La)  &= b_0^{\De} +  b_2^{\De}\,m_\pi^2 + 
\chi_\De\,m_\pi^2\,\ro{log}\,m_\pi/\mu
+ \ca{O}(m_\pi^3)\,.
\end{align}
Here, $\mu$ is an implicit mass scale, chosen here to be $1$ GeV.
 The coefficients
of each polynomial, $b_i^{N,\De}$
 are entirely dependent on the choice of finite-range regularization 
 scheme, and so they are 
regulator-dependent quantities. %and therefore scheme-dependent. 
%In order to achieve an extrapolation
%based on an optimal regularization scale, first
% scale-dependence of the
%low-energy expansion must be removed through renormalization.   
The
renormalization program of FRR combines the scale-dependent $b_i$
coefficients from the chiral loops with the scale-dependent $a_i$
coefficients from the residual series in Eq.~(\ref{eqn:chiral}) 
at each chiral order $i$, such that the 
result is a scale-independent coefficient $c_i$:
\begin{align}
c_0 &= a_0^\La + b_0^{N} + b_0^{\De},\\
c_2 &= a_2^\La + b_2^{N} + b_2^{\De}, \mbox{\,\,etc.}
\end{align}
This means the underlying $a_i$ coefficients undergo a renormalization
from the chiral loop integrals. The renormalized coefficients $c_i$
are an important part of the extrapolation technique. A stable and
robust determination of these parameters
%the success or failure 
%of the renormalization 
%procedure 
forms the heart of determining an optimal scale $\La^\ro{scale}$.
%that can be chosen for the regulator.

The loop integrals can be expressed in a convenient form by taking the
nonrelativistic limit and performing the pole integration for $k_0$.
Renormalization is achieved by subtracting the relevant $b_0^\La$ term 
from the integral, effectively absorbing it into the corresponding 
renormalized coefficient $c_0$. The integrals take the form 
\cite{Wang:2007iw,Wang:2008vb}:
\begin{align}
\label{eqn:tSia}
\tilde{\ca{T}}_N(m_\pi^2\,;\La) 
&=  \f{-\chi_N}{3\pi^2}\int\!\!\ud^3 k\f{k^2u^2(k\,;\La)}
{{(k^2 + m_\pi^2)}^2} - b_0^{N}\,,\\
\label{eqn:tSib}
\tilde{\ca{T}}_\De(m_\pi^2\,;\La)  &= \f{-\chi_\De}{3\pi^2}\int\!\!\ud^3 k\f{k^2 
(2\omega(k) + \Delta)\,u^2(k\,;\La)}{2\omega^3(k)\,[\omega(k) + \Delta]^2} 
- b_0^{\De}\,,
\end{align}
%
%where $\hat{q}$ is the direction of the external momentum introduced 
%by the incoming photon. 
The chiral coefficients $\chi_{N}$ and $\chi_\De$ are 
determined from interactions in 
%constants in terms of 
the chiral Lagrangian of Eq.~(\ref{eqn:lag}):  
%and the relevant Clebsch-Gordan coefficients: %, as summarized by Wang 
%\cite{Wang:2008vb}:
%
\begin{align}
\chi_N^p &= -\f{m_N}{8\pi f_\pi^2}(D+F)^2 = -\chi_N^n,\\
\chi_\De^p &= -\f{m_N}{8\pi f_\pi^2}\f{2\,\ca{C}^2}{9} = -\chi_\De^n.
\end{align}
The chiral expansion of the magnetic moment in Eq.~\ref{eqn:chiral} can 
now be written out in a form renormalized to order $\ca{O}(1)$:
\eqb
\label{eqn:chiralrenorm}
\mu_N^\ro{v} = c_0 + a_2^\La\,m_\pi^2 + \tilde{\ca{T}}_N(m_\pi^2\,;\La) + 
\tilde{\ca{T}}_\De(m_\pi^2\,;\La) + 
 \ca{O}(m_\pi^4)\,.
\eqe

%
%ON THE LATTICE
Since lattice simulations are necessarily carried out on a discrete
spacetime, any extrapolations performed should take into account
finite-volume effects.  $\chi$EFT is ideally suited for
characterizing the leading infrared effects associated with the finite
volume. In order to achieve this, each of the three-dimensional integrals
can be transformed to its form on the lattice using a finite-sum of
discretized momenta, see Allton \textit{et al.}  \cite{Armour:2005mk}
for instance:
\eqb
\int\!\!\ud^3 k \ra \f{{(2\pi)}^3}{L_x L_y L_z} \sum_{k_x,k_y,k_z}. 
\eqe
On the finite-volume lattice, each momentum component is quantized 
in units of $2\pi/L$, that
is $k_i=2\pi\,n_i/L$ for integers $n_i$.  Finite-volume corrections 
$\de^{\ro{FVC}}$
 can be written simply as the difference between the finite sum
and the corresponding integral. It is known that the finite-volume 
corrections saturate to a fixed result for large values of regularization scale 
\cite{AliKhan:2003cu,Hall:2010ai}.
%Following the example set by this article, 
The value $\La'= 2.0$ GeV is chosen
to evaluate all finite-volume corrections independently from the integral cutoff
 scale $\La$ in Eqs.~(\ref{eqn:tSia}) and (\ref{eqn:tSib}). 
This method is equivalent to the more algebraic approach outlined in
Ref.~\cite{Beane:2004tw}.
The finite-volume version of Eq.~(\ref{eqn:chiralrenorm})
 can thus be expressed:
\begin{align}
\label{eqn:finchiral}
\mu_N^\ro{v} &= c_0 + a_2^\La\,m_\pi^2 + (\tilde{\ca{T}}_N(m_\pi^2\,;\La) + 
\de^\ro{FVC}_{N}(m_\pi^2;\La'))\nn\\ 
&+ (\tilde{\ca{T}}_\De(m_\pi^2\,;\La) + \de^\ro{FVC}_{\De}(m_\pi^2;\La')) + 
 \ca{O}(m_\pi^4)\,.
\end{align}

\section{Results}
\label{sect:results}

\subsection{Renormalization flow analysis}
\label{subsect:curves}

In order to obtain the most robust extrapolation, an optimal 
regularization scale is sought. The robustness of the extrapolation 
is characterized by its stability against truncation of the lattice 
data set. That is, a similar extrapolation should be achieved regardless 
of the number of data points used in the fit. 
%In order to obtain the optimal regularization scale, 
%the low-energy coefficient $c_0$ 
%from Eq.~(\ref{eqn:finchiral}) will be calculated across a range 
%of values of $\La$. Thus the renormalization flow can be constructed. 
%Multiple renormalization flow curves may be obtained by constraining the 
%fit window by a maximum, $m_{\pi,\ro{max}}^2$, and sequentially adding  
%points to extend further outside the PCR. 
The optimal regularization scale $\La^\ro{scale}$ may be obtained 
by calculating the low-energy coefficients (e.g. $c_0$) 
from Eq.~(\ref{eqn:finchiral}) for a range of regulator values $\La$. 
Since the lattice simulation results extend outside PCR, the renormalized 
value of the coefficients will be scale dependent. However, the 
analysis in Ref.~\cite{Hall:2010ai} demonstrates that using different 
 amounts of lattice data yields a different scale dependence. 
If the lattice simulation results lie close to the PCR, the scale dependence is 
naturally less than if a more extensive set of lattice data is used. 
The optimal regularization scale is the value of $\La$ at which the same 
value of $c_0$ (or any low-energy coefficient) is obtained regardless 
of the amount of lattice data used. This is the %value of the regularization 
scale where the 
values of the 
low-energy coefficients %are least sensitive to truncation 
%of the lattice results. Furthermore, this optimal scale corresponds to the 
correspond to the values 
%value of regularization scale 
obtained from lattice results within the PCR 
\cite{Hall:2010ai}.

\begin{figure}[bp]
\includegraphics[height=0.950\hsize,angle=90]{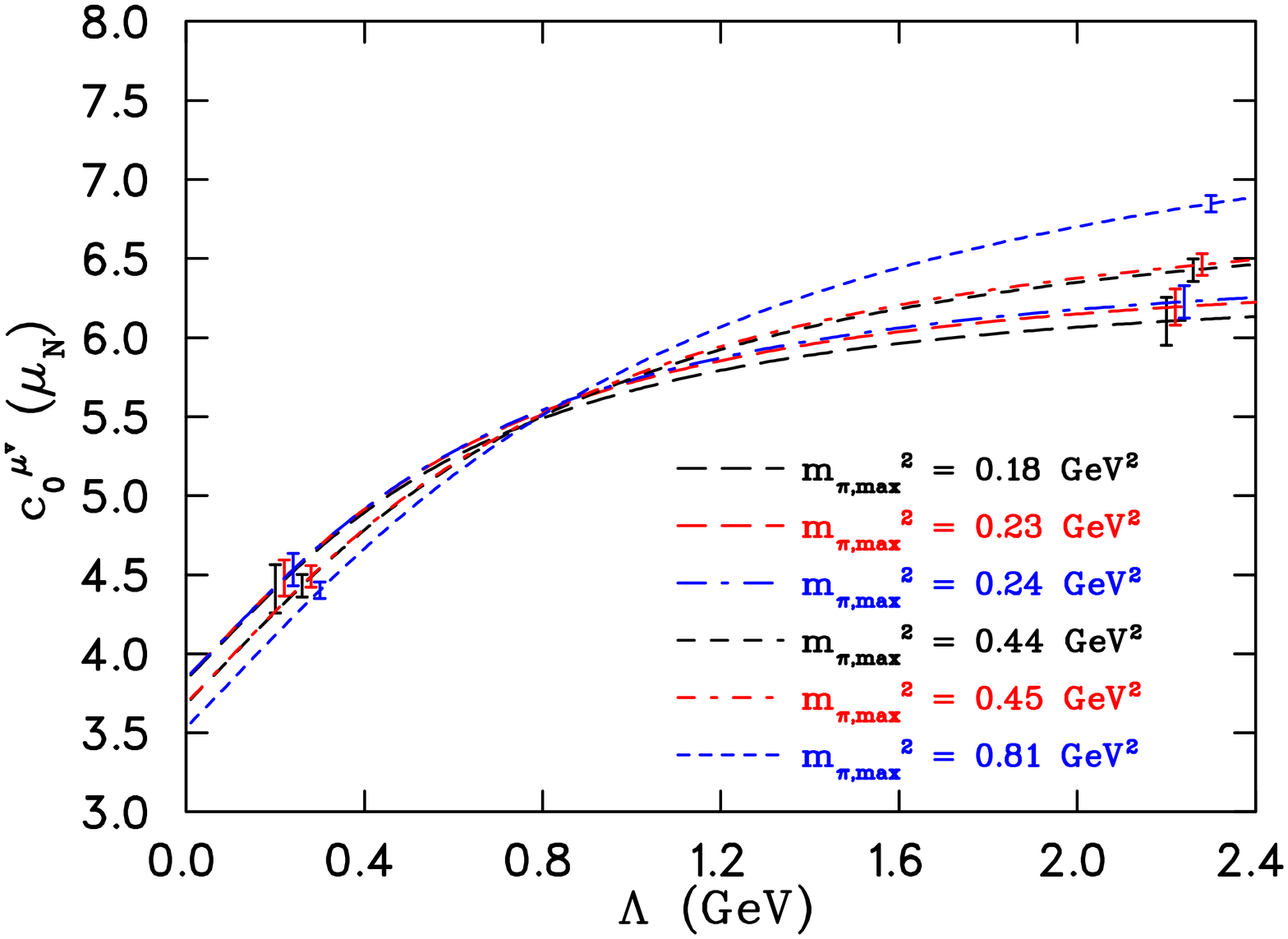}
\vspace{-11pt}
\caption{\footnotesize{(color online). The renormalization flow of
  $c_0$ for $\mu_N^\ro{v}$ obtained using a dipole regulator on 
 QCDSF lattice QCD results. For each curve, two arbitrary values of $\La$ are chosen to indicate the general size of the error bars.}}
\label{fig:c0}
%\end{figure}
\vspace{6mm}
%\begin{figure}[tp]
\includegraphics[height=0.950\hsize,angle=90]{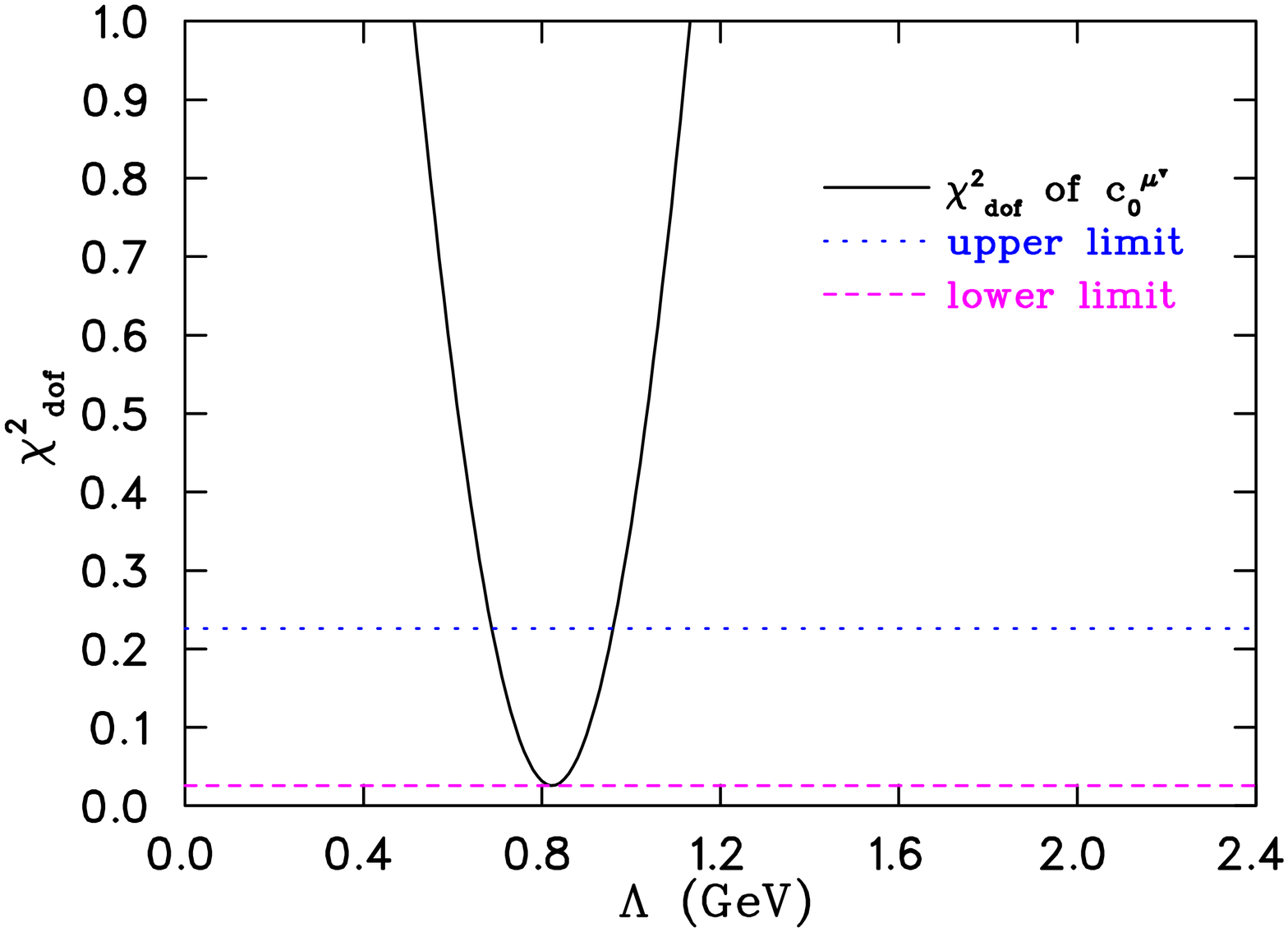}
\vspace{-11pt}
\caption{\footnotesize{(color online). $\chi^2_{dof}$ for 
 the renormalization flow of
  $c_0$ for $\mu_N^\ro{v}$ obtained using a dipole regulator on 
 QCDSF lattice QCD results, up to and including $m_{\pi,\ro{max}}^2 = 0.81$ GeV$^2$.}}
\label{fig:c0chisqdof}
\end{figure}

Consider the behavior of $c_0$ from Eq.~(\ref{eqn:finchiral}) 
as a function of the regularization scale $\La$. Using different 
upper values of $m_{\pi,\ro{max}}^2$, a set of renormalization flow curves 
may be constructed. 
The renormalization flow curves including up to all nine lattice results, 
and using a dipole regulator,  
are plotted on the same set of axes 
in Fig. \ref{fig:c0}.
As more data are included in the fit, a greater degree of regulator dependence 
is observed. Note that there is a reasonably well-defined $\La$ value 
at which the renormalization of $c_0$ is least sensitive to the truncation 
of the data. This indicates that there exists an optimal regularization scale 
embedded 
in the lattice QCD results.

The optimal regularization scale for a dipole  
can be extracted from Fig. \ref{fig:c0} 
using a $\chi^2_{dof}$ analysis. Such an analysis will also provide 
a measure of the systematic uncertainty in the optimal regularization scale.
By plotting $\chi^2_{dof}$ against $\La$, where $dof$ equals 
the number of curves $n$ minus one, a measure of the spread of the 
renormalization 
flow curves can be calculated, and the intersection point obtained.
The $\chi^2_{dof}$ is constructed at each value of $\La$ for 
 $c_0$ (with uncertainty $\de c_0$):
\begin{align}
\chi^2_{dof} = \f{1}{n-1} \sum_{i=1}^{n} \f{{(c_0^i(\La) - \bar{c}_0(\La))}^2}
{{(\de c^i_0(\La))}^2},\\
\bar{c}_0(\La) = \f{\sum_{i=1}^{n}c_0^i(\La)/{{(\de c_0^i(\La))}^2}}
{\sum_{j=1}^{n} 1 / {(\de c_0^j(\La))}^2}.
\end{align}
The indices $i$ and $j$ correspond to data sets with different values of  
$m_{\pi,\ro{max}}^2$.  
The $\chi^2_{dof}$ plot % for a dipole regulator are shown in 
corresponding to Fig. \ref{fig:c0} is shown in Fig. \ref{fig:c0chisqdof}.
%The optimal regularization scale  $\La^\ro{scale}$ 
%is taken to be the central value $\La_\ro{central}$ of the plot, and 
The upper and lower bounds on $\La$ obey 
 the condition $\chi^2_{dof} < \chi^2_{dof, min} + 1/(dof)$. 
Thus the optimal regularization scale for a dipole form 
is: %$\La_{\ro{scale}} = 1.125 - 0.215 +0.195$
%$\La^{\ro{scale}} = 1.13^{+0.22}_{-0.20}$ %(\scriptsize{$\stackrel{+0.22}{-0.20}$}) GeV.
% \scriptsize{$\Big\{\begin{matrix}+0.22\\-0.20 \end{matrix}$}} \normalsize GeV.
%$\Big\{\begin{matrix}+0.22\\-0.20 \end{matrix}$  
$\La^{\ro{scale}} = 0.82^{+0.14}_{-0.13}$ 
 GeV.
This value is %consistent with 
smaller than the optimal regularization scale obtained 
for the nucleon mass using a dipole form \cite{Hall:2010ai} based on 
lattice QCD results from 
JLQCD \cite{Ohki:2008ff}, PACS-CS \cite{Aoki:2008sm} and 
CP-PACS \cite{AliKhan:2001tx}. Nevertheless, the value of the optimal scale 
is the same order of magnitude as those calculated from the nucleon mass 
analyses, and previous studies of the magnetic moment \cite{Leinweber:1998ej}. 
In addition, it has been assumed that all available lattice results 
should be used in constructing the fit. This is not necessarily the case. 
 The method described in Ref.~\cite{Hall:2011en} 
outlines a procedure for determining the 
optimal fit window of pion masses, as discussed in the following section. 
%This is strong evidence that,
%for a given functional form, the optimal 
%regularization scale 
%is associated with an intrinsic scale, which characterizes the size of 
%the nucleon, as probed by the pion.

%

\subsection{Chiral extrapolations}
\label{subsect:extrap}

Using the optimal regularization scale, 
extrapolations can be made on various lattice volumes. 
In order to determine the most suitable number of data points to be used 
for fitting the lattice results, %the method described in 
%Ref. \cite{Hall:2011en} is used. 
one may perform several extrapolations, using a varying number of points 
each time, and compare the size of the statistical and systematic error 
estimates. The statistical uncertainty comprises contributions from the fit 
coefficients, 
and the optimal regularization scale $\La^{\ro{scale}}$, which is assumed to 
be independent of the other fit coefficients, and its contribution is  
added in quadrature.  
The axial coupling $g_A$ 
and the pion decay constant $f_\pi$ are assumed to be sufficiently 
well determined experimentally. 
The systematic uncertainty in the extrapolation 
is estimated by comparing the results from different regulator functional 
forms. 
%Since the finite-volume corrections due to a sharp cutoff regulator are 
%poorly convergent, a fair approximation is 
The triple-dipole regulator introduced in Refs.~\cite{Hall:2010ai,Hall:2011en} 
is considered, which interpolates 
between the dipole and the sharp cutoff regulators. 
The resultant chiral extrapolation using a dipole regulator is compared 
to that using a triple-dipole regulator in order to estimate the 
systematic uncertainty. 

\begin{figure}[tp]
\includegraphics[height=0.95\hsize,angle=90]{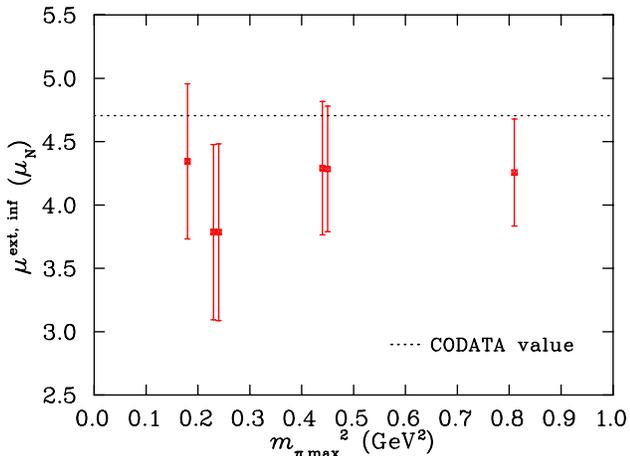}
\vspace{-11pt}
\caption{\footnotesize{(color online). Behaviour of the extrapolation of $\mu_N^\ro{v}$ to the physical point vs $m_{\pi,\ro{max}}^2$. In each case, the value 
of $\La^{\ro{scale}}$ %$\La_{\ro{central}}$ 
is used, %(for a dipole regulator)
 as obtained from the corresponding $\chi^2_{dof}$ analysis. The error bars include the statistical and systematic uncertainties %in $c_0$ 
added in quadrature. %The smallest uncertainty corresponds to a 
%maximum pion mass of ${m}_{\pi,\ro{max}}^2 = 0.81$ GeV$^2$.
}}
\label{fig:muvsmpisq}
\vspace{-2mm}
\end{figure}

\begin{figure}
\includegraphics[height=0.95\hsize,angle=90]{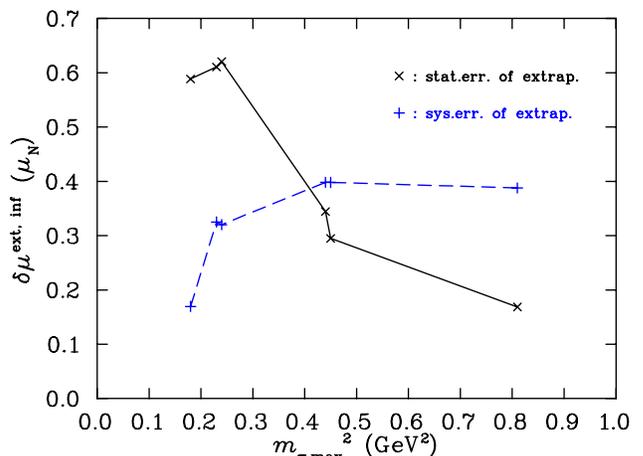}
\vspace{-11pt}
\caption{\footnotesize{(color online). Magnitude of the statistical and systematic error bar in the extrapolation of $\mu_N^\ro{v}$ to the physical point vs $m_{\pi,\ro{max}}^2$. In each case, the value of $\La^{\ro{scale}}$ %scale $\La_{\ro{central}}$ 
is used, %(for a dipole regulator) 
as obtained from the corresponding $\chi^2_{dof}$ analysis.  At a maximum pion mass of $\bar{m}^2 = {m}_{\pi,\ro{max}}^2 = 0.44$ GeV$^2$, the best compromise between statistical and systematic uncertainty is achieved.}}
\label{fig:muvsmpisqss}
\end{figure}

The quadrature sum of the statistical and systematic uncertainties in the 
extrapolation of $\mu_N^\ro{v}$ to the physical point, for different values of 
$m_{\pi,\ro{max}}^2$, is shown in
Fig. \ref{fig:muvsmpisq}.
 Ideally, 
one should find a best value of the upper limit $m_{\pi,\ro{max}}^2$, as indicated 
by the best compromise between statistical and systematic effects. 
Fig. \ref{fig:muvsmpisq} indicates that the smallest error bar occurs 
when all nine lattice points are included. 
This set corresponds to a maximum pion-mass value: $m_{\pi,\ro{max}}^2 = 0.81$
GeV$^2$. 
However, it is helpful to know the relative contributions from 
statistical and systematic sources. Table~\ref{table:ss} summarizes the 
breakdown of each error bar into its sources. 
Fig. \ref{fig:muvsmpisqss} shows the magnitude of the statistical and 
systematic error bars for different values of $m_{\pi,\ro{max}}^2$. 
Clearly, the general trend of the statistical error bar decreases as more 
lattice results are considered, and likewise the general trend of the 
systematic error bar increases. 
At $m_{\pi,\ro{max}}^2 = 0.44$ GeV$^2$, the statistical and systematic 
error bars are closest in magnitude, which indicates the proximity of a 
`sweet spot' (denoted $m_{\pi,\ro{max}}^2 = \bar{m}^2$), 
at which the best trade-off between statistical and 
systematic uncertainty is achieved.
%That is, using the lightest seven lattice points, the best 
%compromise between statistical and systematic uncertainty is achieved for this 
%set of lattice results.

%Note that the extrapolated value of the magnetic moment differs from 
%the CODATA \cite{Mohr:2008fa} / PDG \cite{Nakamura:2010zzi}
% value by approximately three standard deviations.

%\vspace{6mm}

\begin{table}[tp]
  \caption{\footnotesize{Results for the isovector nucleon magnetic moment for different values of $m_{\pi,\ro{max}}^2$, extrapolated to the physical point, corresponding to Fig. \ref{fig:muvsmpisq}. The uncertainty in $\mu_N^\ro{v}(m_{\pi,\ro{phys}}^2)$ is provided in the following order: the statistical uncertainty, %due to the fit coefficients, 
the optimal regularization scale $\La^{\ro{scale}}$, and the systematic uncertainty due to the regulator functional form, respectively.   
}}
\vspace{-6pt}
  \newcommand\T{\rule{0pt}{2.8ex}}
  \newcommand\B{\rule[-1.4ex]{0pt}{0pt}}
  \begin{center}
    \begin{tabular}{ll}
      \hline
      \hline
      \T\B            
      $m_{\pi,\ro{max}}^2$(GeV$^2$)  &  \qquad $\mu_N^\ro{v}(m_{\pi,\ro{phys}}^2)$ ($\mu_N$)\\
      \hline     
      $0.185$   &\T\qquad $4.35(13)(57)(17)$\\
      $0.230$   &\T\qquad $3.79(10)(60)(33)$\\
      $0.240$   &\T\qquad $3.79(9)(61)(32)$\\
      $0.436$   &\T\qquad $4.29(7)(34)(40)$\\
      $0.449$   &\T\qquad $4.29(7)(29)(40)$\\
      $0.810$   &\T\qquad $4.26(5)(16)(39)$\\
      \hline
    \end{tabular}    
  \end{center}
  \label{table:ss}
\vspace{-3mm}
\end{table}

\begin{figure}[tp]
\vspace{2mm}
\includegraphics[height=0.950\hsize,angle=90]{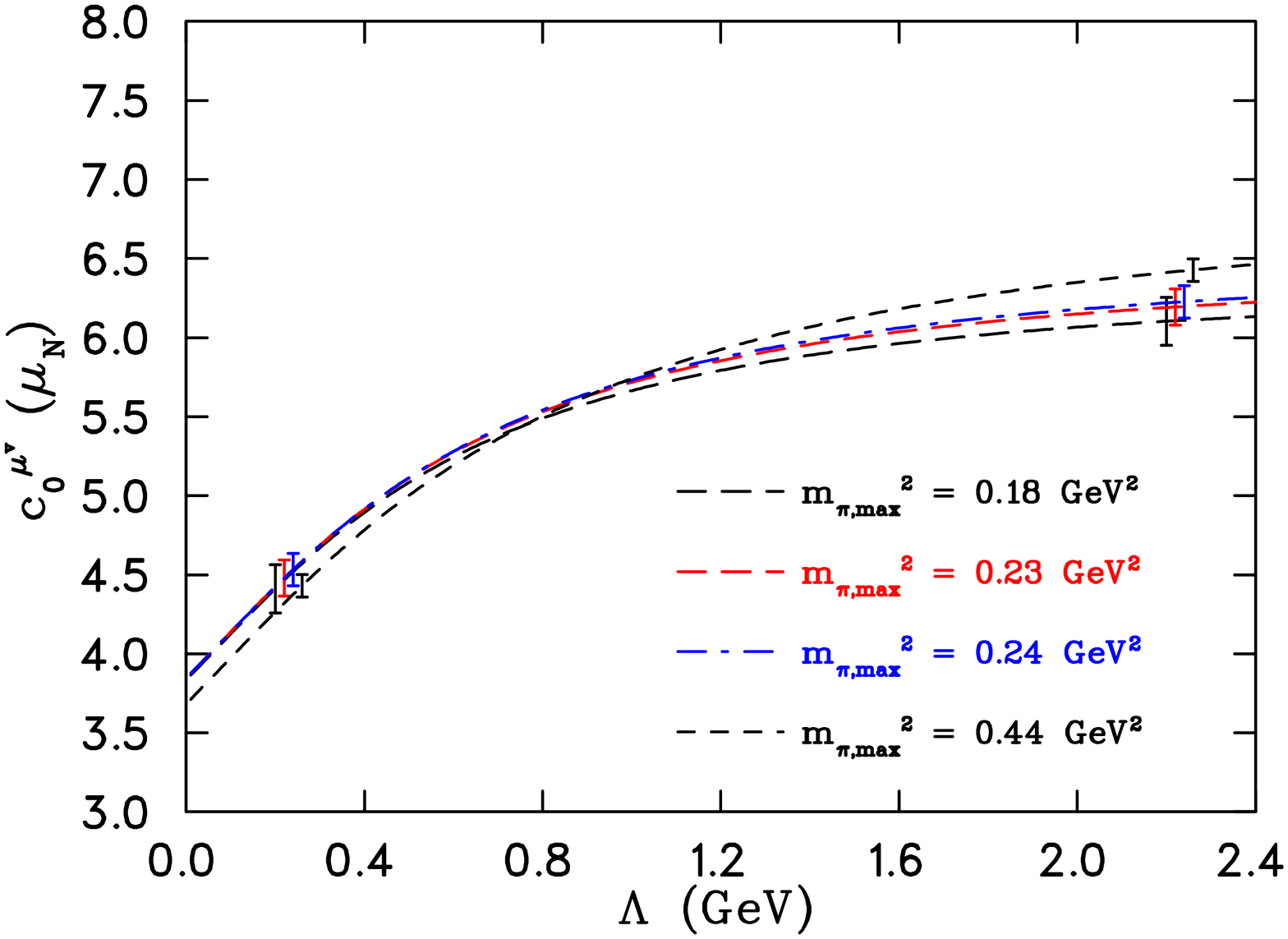}
\vspace{-11pt}
\caption{\footnotesize{(color online). The renormalization flow of
  $c_0$ for $\mu_N^\ro{v}$ obtained using a dipole regulator on 
 QCDSF lattice QCD results. Only the lightest seven lattice results are used. For each curve, two arbitrary values of $\La$ are chosen to indicate the general size of the error bars.}}
\label{fig:c0_7p}
%\end{figure}
\vspace{6mm}
%\begin{figure}[tp]
\includegraphics[height=0.950\hsize,angle=90]{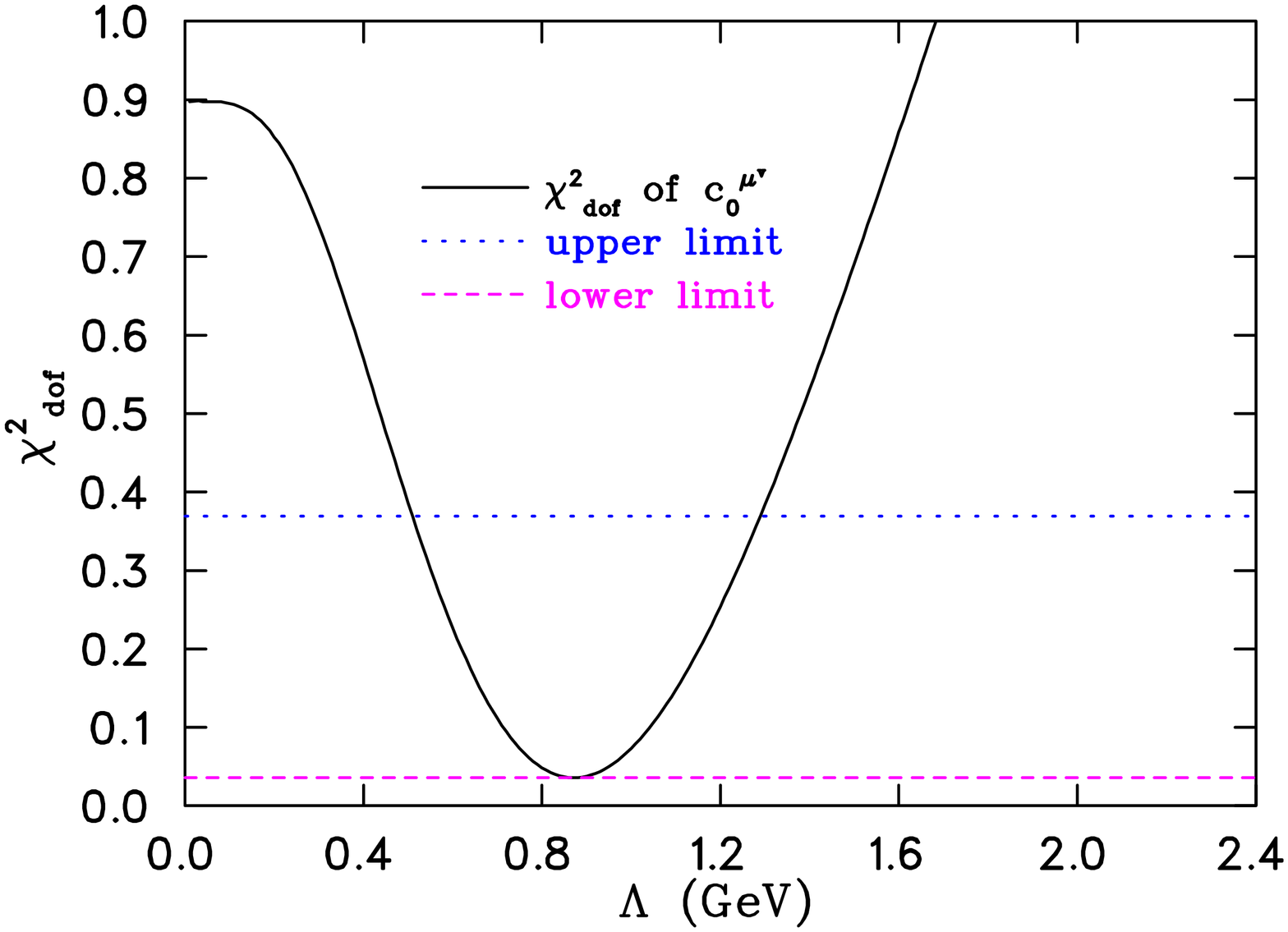}
\vspace{-11pt}
\caption{\footnotesize{(color online). $\chi^2_{dof}$ for 
 the renormalization flow of
  $c_0$ for $\mu_N^\ro{v}$ obtained using a dipole regulator on 
 QCDSF lattice QCD results, up to and including $m_{\pi,\ro{max}}^2 = 0.44$ GeV$^2$ only.}}
\label{fig:c0chisqdof_7p}
\end{figure}

The renormalization flow curves corresponding at most to a value of  
$m_{\pi,\ro{max}}^2 = 0.44$ GeV$^2$ are shown in Fig. \ref{fig:c0_7p}, and the 
corresponding $\chi^2_{dof}$ plot is shown in Fig. \ref{fig:c0chisqdof_7p}. 
Clearly, the statistical contribution to the uncertainty in the extrapolation 
is larger than in the case where all lattice results are used. 
This is reflected in the larger uncertainty in the identification 
of the optimal regularization scale (at optimal $m_{\pi,\ro{max}}^2 = \bar{m}^2$:  
$\La^{\ro{scale}}_{\bar{m}^2} = 0.87^{+0.42}_{-0.36}$. 
This value is consistent with the 
optimal regularization scale obtained 
for the nucleon mass, using a dipole form \cite{Hall:2010ai}. This  
provides evidence for the successful extraction 
of the intrinsic scale in the nucleon-pion 
interaction.
Since this value of $\La^{\ro{scale}}_{\bar{m}^2}$ 
is obtained from the best compromise 
between statistical and systematic uncertainty, it  will be used in 
the following chiral extrapolation of the isovector nucleon magnetic moment.  
Only the lightest seven lattice points 
(corresponding to $m_{\pi,\ro{max}}^2 = 0.44$
GeV$^2$) will be used in the fit.

\begin{figure}[bp]
\includegraphics[height=0.950\hsize,angle=90]{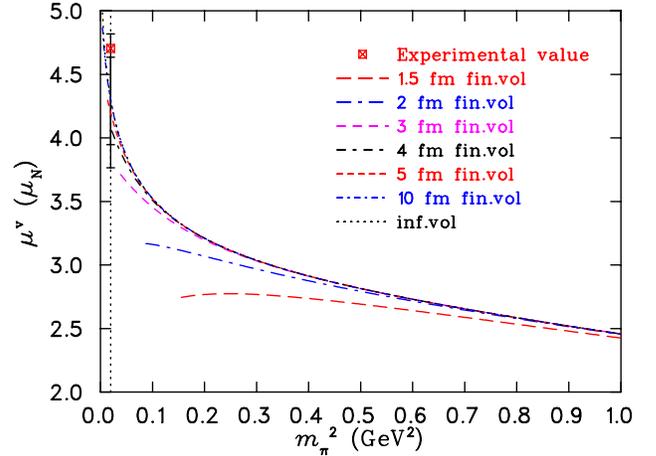}
\vspace{-11pt}
\caption{\footnotesize{(color online). Extrapolations of $\mu_N^\ro{v}$ 
at different finite volumes and infinite volume. % using a dipole regulator with
%$\La = \La^{\ro{scale}}_{\bar{m}^2}$. 
 The curves are 
based on lattice QCD results from QCDSF, lattice sizes: $1.7-2.9$ fm. 
The experimental value 
is  marked \cite{Mohr:2008fa,Nakamura:2010zzi}. 
%Only the lightest seven points are used in the fit, corresponding to a value of $m_{\pi,\ro{max}}^2 = 0.44$ GeV$^2$, and 
In all finite-volume extrapolations, the provisional constraint $m_\pi L > 3$ 
is used. 
%An estimate in the uncertainty in the extrapolation due to 
%$\La^{\ro{scale}}_{\bar{m}^2}$ 
% has been calculated from Fig. \ref{fig:c0chisqdof_7p}, and is indicated at the physical value of $m_\pi^2$.
}}
\label{fig:extrapbare}
\end{figure}

 Consider the behaviour of the magnetic moment as a function of the quark mass.
%Extrapolation curves 
%are then plotted for infinite-volume and a variety of finite-volumes at 
%which current lattice QCD results are produced. 
%
 Extrapolation curves corresponding to 
infinite volume, and a variety of finite volumes are shown in 
Fig. \ref{fig:extrapbare}. 
For each curve, only the values for which $m_\pi L > 3$ are plotted. 
These finite volumes include typical values at  
which current lattice QCD results are produced. 
For example, a full QCD simulation at physical quark masses on a 
($4$ fm)$^3$ volume will significantly underestimate the infinite-volume 
result. 
%
%Extrapolation curves 
%are then plotted for infinite-volume and a variety of finite-volumes at 
%which current lattice QCD results are produced. 
 % 
% provisionally, to avoid undesired effects of the $\epsilon$-regime.
%
These curves indicate that a box length of $L>5\,\,{\rm fm}$  is required 
to achieve an extrapolation within $2\%$ of the infinite-volume 
result. 
%The infinite-volume extrapolation to the physical point 
%is within $2\%$ of the experimentally 
%derived value. 

\begin{figure}
\includegraphics[height=0.950\hsize,angle=90]{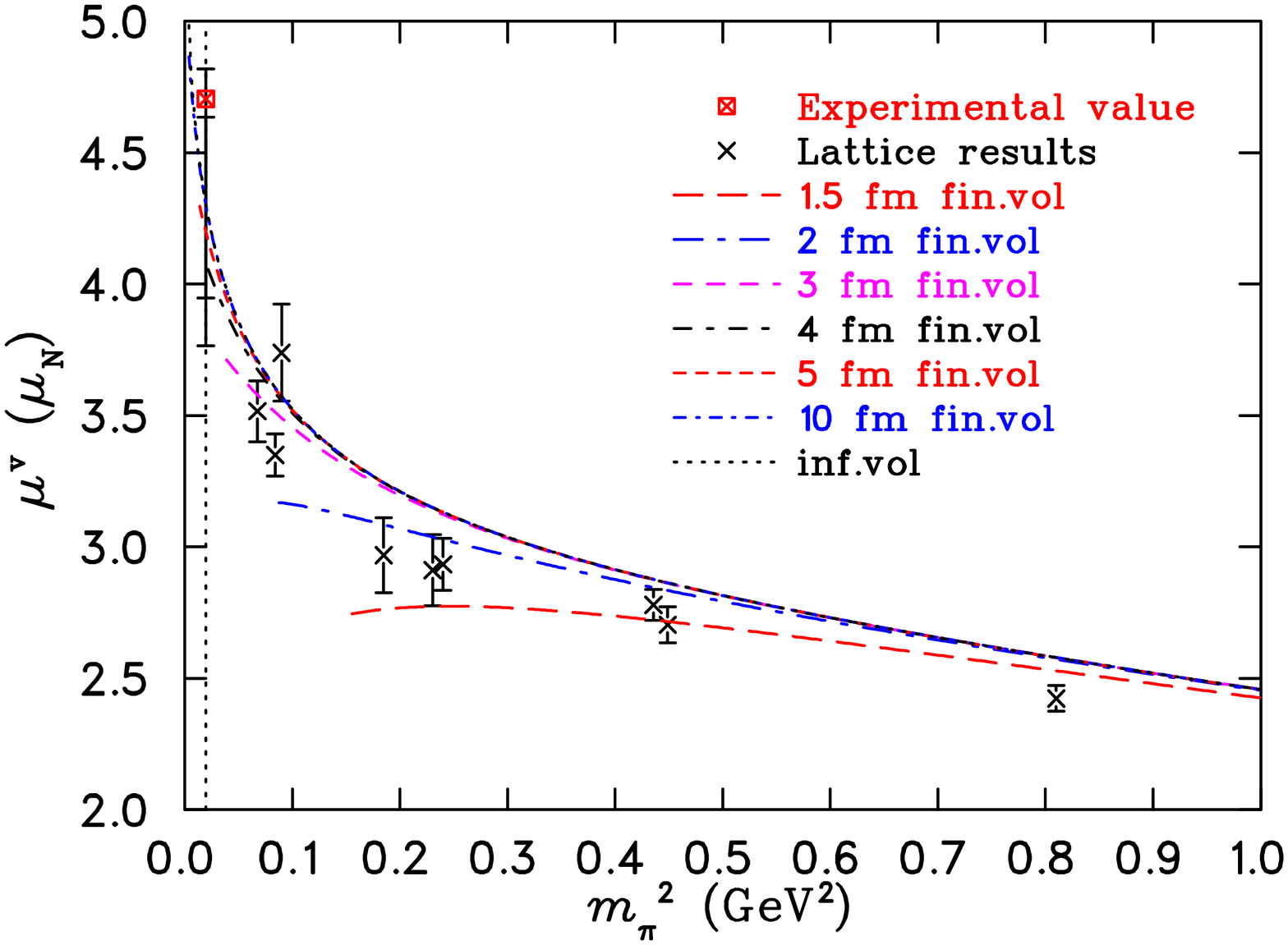}
\vspace{-11pt}
\caption{\footnotesize{(color online). Extrapolations of $\mu_N^\ro{v}$ 
at different finite volumes and infinite volume. % using a dipole regulator with
%$\La = \La^{\ro{scale}}_{\bar{m}^2}$. 
The finite-volume lattice QCD results from 
Ref.~\cite{Collins:2011mk} 
are plotted for comparison, with box sizes in the range $1.7-2.9$ fm.  
%The provisional constraint $m_\pi L > 3$ 
%is used.
In all finite-volume extrapolations, the provisional constraint $m_\pi L > 3$ 
is used. 
%Only the lightest seven points are used in the fit, corresponding to a value of $m_{\pi,\ro{max}}^2 = 0.44$ GeV$^2$, and in all finite-volume extrapolations, the provisional constraint $m_\pi L > 3$ 
%is used. 
%An estimate in the uncertainty in the extrapolation due to 
%$\La^{\ro{scale}}_{\bar{m}^2}$ 
% has been calculated from Fig. \ref{fig:c0chisqdof_7p}, and is indicated at the physical value of $m_\pi^2$.
}}
\label{fig:extrapfin}
\includegraphics[height=0.950\hsize,angle=90]{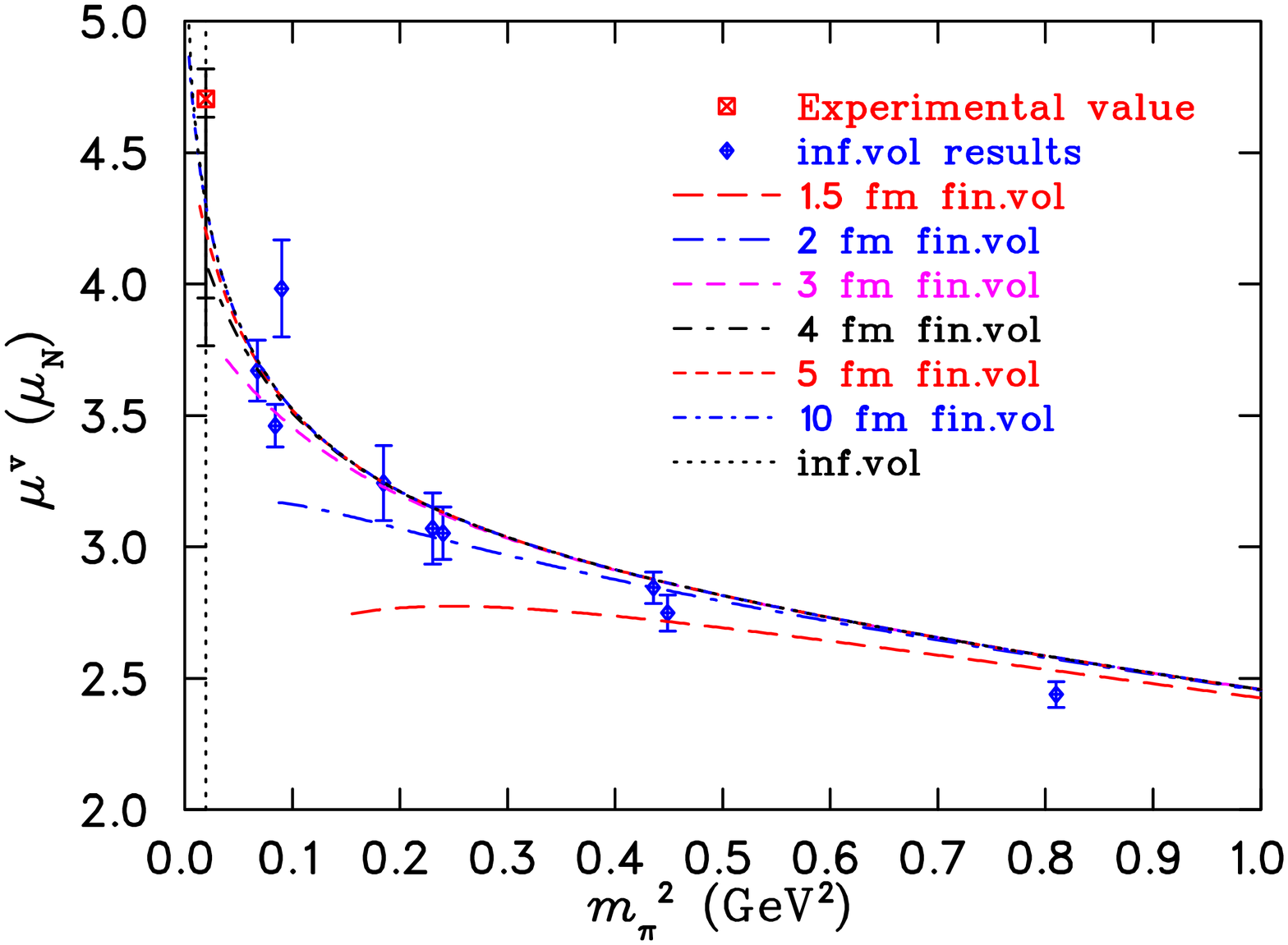}
\vspace{-11pt}
\caption{\footnotesize{(color online). Extrapolations of $\mu_N^\ro{v}$ 
at different finite volumes and infinite volume.% using a dipole regulator with 
%$\La = \La^{\ro{scale}}_{\bar{m}^2}$. 
The lattice QCD results displayed have been 
corrected to infinite volume.  
Only the lightest seven points are used in the fit, corresponding to a value of $m_{\pi,\ro{max}}^2 = 0.44$ GeV$^2$. % and in all finite-volume extrapolations, the provisional constraint 
%$m_\pi L > 3$ 
%is used. 
%In all finite-volume extrapolations, the provisional constraint $m_\pi L > 3$ 
%is used. 
%An estimate in the uncertainty in the extrapolation due to 
%$\La^{\ro{scale}}_{\bar{m}^2}$ 
% has been calculated from Fig. \ref{fig:c0chisqdof_7p}, and is indicated at the physical value of $m_\pi^2$.
}}
\label{fig:extrapinf}
\end{figure}

The finite-volume expansion of Eq.~(\ref{eqn:finchiral}) 
is constrained 
by the lattice simulation results from several different volumes in the range 
$1.7-2.9$ fm, as shown in 
 Fig. \ref{fig:extrapfin}. 
The infinite-volume extrapolation is fit to the lattice simulation results 
only after the results have been corrected to infinite volume. These points 
are shown in Fig. \ref{fig:extrapinf}. 
The extrapolation to the physical point also includes an 
inner error bar representing only the statistical uncertainty, and 
an outer error bar, which also includes the systematic uncertainty due to the 
regulator in quadrature. 

In all extrapolations, the strange quark loops have been unquenched, and the 
effects of kaons loops that would occur in an $\SU(3)$ lattice calculation 
are estimated.  The result is a change of only $\approx 0.7$\% larger 
at the physical point when kaons loops are included. 
 
The finite-volume extrapolations of Fig. \ref{fig:extrapbare} 
are generally useful for estimating the 
result of a lattice QCD calculation at certain box sizes. This can also 
provide a benchmark for estimating the outcome of a lattice QCD simulation 
at larger and untested box sizes.
%Note that even a relatively standard 
%$3$ fm lattice box length will differ significantly from the experimental 
%value at the physical point.
%Since the data points in Fig. \ref{fig:extraps} are at differing finite 
%volumes, the infinite-volume corrected data points are displayed 
% in Fig. \ref{fig:extrapsinf}.

%
%\vspace{6mm}

\newpage
\section{Conclusion}
\label{sect:conc}
\vspace{-2mm}
The technique for obtaining an optimal regularization scale 
from lattice QCD results   
has been investigated in 
the context of the magnetic moment of the isovector nucleon,  using 
recent precision lattice QCD results from QCDSF. An optimal 
regularization scale 
%for a dipole regulator was obtained. This was achieved 
 was identified 
by analyzing the 
renormalization flow of the low-energy coefficient $c_0$ with respect 
to the scale $\La$, 
whilst extending beyond the power-counting regime.  
 An optimal value of $m_{\pi,\ro{max}}^2$ was also obtained, where the 
statistical and systematic error estimates of a chiral extrapolation 
are comparable in magnitude. This value $\bar{m}^2$ provides 
a guide to the range of pion masses in which finite-range regularization 
 techniques are not the dominant source of uncertainty in a chiral 
extrapolation. 

%\vspace{-0.0mm}

A regularization scale $\La^\ro{scale}_{\bar{m}^2}$ was determined where the 
renormalization of $c_0$ is least sensitive to the truncation of 
the lattice QCD results.
The value of the optimal regularization scale 
was consistent with results from the 
nucleon mass analysis. 
Thus an intrinsic scale has been uncovered, which characterizes 
the energy scale of the nucleon-pion interaction.
The result therefore further demonstrates the success of the procedure for 
using lattice QCD results
 to extrapolate an observable to the low-energy region of QCD. 

Using the value of the intrinsic scale, the extrapolation of
 the magnetic moment to the physical pion mass and infinite-volume 
 lattice box size is consistent with the experimental value. 
More importantly, the finite-volume 
extrapolations provide a benchmark for estimating the outcome of a 
lattice QCD simulation at realistic or optimistic lattice sizes.
This serves to emphasize the importance of achieving 
large volumes in realizing the correct nonanalytic behaviour 
found in nature. 
\vspace{-5mm}
\begin{acknowledgments}
\vspace{-2mm}
We would like the thank James Zanotti for many helpful discussions. 
This research 
is supported by the Australian Research Council. % through Grant DP110101265. 
\end{acknowledgments}

%BIBLIOGRAPHY
%\bibliographystyle{apsrev4-1}
\bibliographystyle{apsrev} 
\vspace{-5mm}
\bibliography{magref}

%*****************************************%

\end{document}